\documentclass[12pt,preprint]{aastex}
\shortauthors{JIM\'ENEZ-VICENTE ET AL.}

\begin{document}

%\bibliographystyle{aastex}

%%%%%%%% Title %%%%%%%

\title {Probing the dark matter radial profile in lens galaxies and the size of X-ray emitting region in quasars with microlensing}
%Dark Matter Fraction in Lensed Galaxies and Quasar Structure from X-ray and Optical Microlensing

%%%%%% Author list  %%%%%%%%%%%%%%%%%
\author{J. JIM\'ENEZ-VICENTE\altaffilmark{1,2},  E. MEDIAVILLA\altaffilmark{3,4}
,
C. S. KOCHANEK\altaffilmark{5},  J. A. MU\~NOZ\altaffilmark{6,7}}

\altaffiltext{1}{Departamento de F\'{\i}sica Te\'orica y del Cosmos, Universidad
 de Granada, Campus de Fuentenueva, 18071 Granada, Spain}
\altaffiltext{2}{Instituto Carlos I de F\'{\i}sica Te\'orica y Computacional, Un
iversidad de Granada, 18071 Granada, Spain}
\altaffiltext{3}{Instituto de Astrof\'{\i}sica de Canarias, V\'{\i}a L\'actea S/
N, La Laguna 38200, Tenerife, Spain}
\altaffiltext{4}{Departamento de Astrof\'{\i}sica, Universidad de la Laguna, La 
Laguna 38200, Tenerife, Spain}
\altaffiltext{5}{Department of Astronomy and the Center for Cosmology and Astroparticle Physics, The Ohio State University, 4055 McPherson Lab, 140 West 18th Avenue, Columbus, OH, 43221 }
\altaffiltext{6}{Departamento de Astronom\'{\i}a y Astrof\'{\i}sica, Universidad
 de Valencia, 46100 Burjassot, Valencia, Spain.}
\altaffiltext{7}{Observatorio Astron\'omico, Universidad de Valencia, E-46980 Paterna, Valencia, Spain}

%%%%%%%%%%%%%%%% Abstract %%%%%%%%%%%%%%%%%%%%%%%%%%%%

\begin{abstract}

We use X-ray and optical microlensing measurements 
to study the shape of the dark matter density profile in the lens galaxies and the size of the (soft) X-ray emission region. 
We show that single epoch X-ray microlensing is sensitive to the source size. Our results, in good agreement with previous estimates, show that the size of the X-ray emission region scales roughly linearly with the black hole mass, with a half light radius of  $R_{1/2}\simeq(24\pm14) r_g$ where $r_g=GM_{BH}/c^2$. This corresponds to a size of  $\log(R_{1/2}/cm)=15.6^{+0.3}_{-0.3}$ or $\sim$ 1 light day for a black hole mass of $M_{BH}=10^9 M_\sun$. We simultaneously estimated the fraction of the local surface mass density in stars, finding that the stellar mass fraction is $\alpha=0.20\pm0.05$ at an average radius of $\sim 1.9 R_{e}$, where $R_e$ is the effective radius of the lens. This stellar mass fraction is insensitive to the X-ray source size and in excellent agreement with our earlier results based on optical data.
By combining X-ray and optical microlensing data, 
we can divide this larger sample into two radial bins. We find that the surface mass density in the form of stars is
$\alpha=0.31\pm0.15$ and $\alpha=0.13\pm0.05$ at $(1.3\pm0.3) R_{e}$ and $(2.3\pm0.3) R_{e}$, respectively, in good agreement with expectations and some previous results.

\end{abstract}


\section{Introduction}\label{sec1}
The abundance of dark matter in galaxies and the structure of the central engines of quasars are two interesting astrophysical problems that can be probed by gravitationally lensed quasars. The presence of point-like stars and their remnants in the otherwise smooth distribution of matter of the lens galaxy induces very strong local changes in the gravitational field that give rise to large changes in the magnification
of the lensed source compared to a smooth model known as microlensing (see the review by Wambsganss 2006). The amplitude of these anomalies is sensitive to the local stellar surface mass density fraction as compared to that of the dark matter at the image location (Schechter \& Wambsganss 2002). It is also sensitive to the source size, because larger sources more heavily smooth the magnification patterns, and so are less magnified than smaller sources. Both effects are important for the amplitude and statistics of the microlensing we see in the images of multiply imaged quasars.

Two basic experimental approaches have been used to measure microlensing in lensed quasars: photometric monitoring and single epoch spectroscopy/photometry. Photometric monitoring measures the magnification changes (microlensing variability) induced by the relative motions of the quasar source, the lens galaxy stars and the observer (Chang \& Refsdal 1979; Gott 1981; Kayser, Refsdal \& Stabell 1986) by comparing the light curves of the lensed images. After correcting the curves for the time delay between the images, 
the time varying microlensing signal can be analyzed (e.g. Kochanek 2004). Alternatively, if frequent monitoring of the source is not available, valuable information can still be extracted by using single epoch spectroscopy/photometry. In this case, emission lines 
or a smooth macrolens model can be used
as a reference with respect to which the microlensing of the different images can be measured. 
The advantage of using the emission lines as a reference is that they are much less sensitive to microlensing (e.g. Guerras et al. 2013)
and systematic errors in the macro lens model (see Mediavilla et al. 2009, hereafter MED09).
Single epoch microlensing magnification estimates are observationally much less expensive than photometric monitoring, and can be easily obtained for relatively large samples of lensed quasars. The challenge is to adequately control for the systematic uncertainties in the reference magnification created by time variability, lens substructures and absorption/extinction. 

Gravitational microlensing in the optical has proven a very powerful tool in many studies of individual lenses to estimate the size
of the quasar accretion disks (see, e.g., Morgan et al. 2008, 2010, 2012, Mediavilla et al. 2011b, Mu\~noz et al. 2011, Motta et al. 2012, Mosquera et al. 2009, 2011, 2013, Rojas et al. 2014) and also the fraction of surface mass density in the form of stars (Schechter \& Wambsganss 2004, Kochanek et al. 2004, 2006, Bate et al. 2011).
In the optical, most of the works based on large lens samples are focused on either the fraction of mass in stars $\alpha$ (e.g., Mediavilla et al. 2009) or the quasar size $r_s$ (e.g. Blackburne et al. 2011; Jim\'enez-Vicente et al. 2012) separately. However, since microlensing is sensitive to both physical effects, some degeneracy was expected in the microlensing based estimates of these parameters (see MED09). This has been shown by Jim\'enez-Vicente et al. (2015) who, despite the strong covariance between stellar mass fraction $\alpha$ and source size $R_{1/2}$, found reasonably good estimates for both parameters of $\alpha= 0.21\pm0.14$ (at a radius of approximately $1.8 R_{e}$ where $R_e$ is the effective radius of the lens, and $R_{1/2}=2.0^{+1.0}_{-0.6}\times 10^{16}\sqrt{M/0.3M_\sun}$  cm at an average rest wavelength of $\lambda=1734$ \AA, where $M$ is the mean mass of the microlenses, .

Microlensing of multiply imaged quasars has also been observed in X-rays (e.g. Pooley et al. 2006, 2007, Blackburne et al. 2006, Morgan et al. 2008, Chartas et al. 2009, Dai et al. 2010, Morgan et al. 2012, Blackburne et al., 2014, 2015, Mosquera et al. 2013, MacLeod \& Morgan, 2014). Size estimates in these studies generally find that the soft X-rays are emitted from a region with a size of $R_{1/2}\sim 20r_g$ (where $r_g=GM_{BH}/c^2$ is the gravitational radius of the black hole), while hard X-rays may come from a slightly more compact region. The ratio of the sizes of the optical and the soft X-ray sources is of order $\sim$ 10.
Studies using samples of lenses with X-ray data have been used to estimate the stellar mass fraction $\alpha$ (Pooley et al. 2012, Schechter et al. 2014), and they have generally assumed that the size of the X-ray emitting region is small enough to have little impact on the estimate of $\alpha$. MED09 showed that there was a significant covariance between stellar mass fraction $\alpha$ and the source size $r_s$ in the optical, and Jim\'enez-Vicente et al. (2015) found that despite the strong covariance, a determination of both parameters was still possible.
As X-ray microlensing has proven to be sensitive to both size and fraction of mass in stellar mass objects, it is reasonable to wonder
whether there is also a covariance between these two parameters for X-rays, similar to what is found at optical wavelengths, and how it may affect the estimates of both parameters. A joint study of both parameters using X-ray microlensing estimates is therefore needed to clarify this point.

Beyond measuring a mean value for the stellar mass fraction at an average radius, $\alpha$, measuring  the radial profile 
$\alpha$ is a key ingredient in understanding how galaxies formed; in particular the interaction of dark and baryonic matter during the initial collapse (including processes like baryonic cooling, settling, star formation and feedback) and subsequent mergers (cf. Diemand \& Moore 2011). Most studies addressing this issue use other methods (X-ray emission, dynamics or strong lensing) to estimate total masses, and use the brightness/color distribution and a suitable IMF to estimate the stellar mass (see, for example, the review by Courteau et al. 2014). But measuring radial profiles using these procedures is model dependent, particularly through the IMF, which is itself a subject of study using lensing (e.g. Treu et al. 2010, Sonnenfeld et al. 2012).
Microlensing can provide an estimate of the local stellar mass fraction $\alpha$ at the location of the images without any strong dependence on the specific shape of the mass distribution.
In this respect, microlensing studies, either based on X-ray or optical data, have mainly focused on obtaining an estimate of the fraction of mass in stars at an average distance. Pooley et al. (2012) made a first attempt to measure a radial gradient in the stellar fraction and did not detect a significant gradient. 

Here we make a new attempt to detect such a radial gradient in the stellar mass fraction. 
We will combine the available optical and X-ray data to estimate the fraction of mass in the form of stars at different radii within lens galaxies. We will subsequently check the consistency of our estimates of the stellar/dark matter fraction by comparing them with previous results and models. In Section 2 we analyze the sensitivity of X-ray microlensing to the source size, 
and we discuss the dependence of X-ray sizes on the black hole mass. In Section 3, we address the
joint estimate of the stellar mass fraction and the typical size of the X-ray emitting region. 
Section 4 is devoted to study the radial profile of the stellar mass fraction in lens galaxies. Finally, the main results are summarized in Section 5.

\section{The Dependence of X-ray Sizes on Black-Hole Mass}\label{sec2}\

We start by estimating the size of the X-ray emitting region for 10 quadruple lens systems using the flux ratios from Schechter et al. (2014), who used the soft (0.5-8 keV) X-rays fluxes from Pooley et al. (2007) and Blackburne et al. (2011). To compare microlensing magnification estimates for different models to observations, we follow the procedures of Jim\'enez-Vicente et al. (2012).
We compute magnification maps for each of the four images in the 10 systems using the Inverse Polygon Mapping technique (Mediavilla et al. 2006, 2011a). We take the values for $\kappa$ and $\gamma$ provided by Schechter et al. (2014), and put 20\% of the surface mass density in the form of stars, as derived from microlensing in the optical by Jim\'enez-Vicente et al. (2015). We use stars of fixed mass, which we have chosen to be $M=0.3M_\sun$. The maps have $2000\times 2000$ pixels and span 100 light days with a fixed pixel size of 0.05 light days. 

\begin{deluxetable}{llr}
\tabletypesize{\footnotesize}
\tablewidth{0pt}
%\rotate
\tablecolumns{3}
\tablecaption{Microlensing data.\label{tab1}}
\tablehead{\multicolumn{3}{c}{X-rays}\\ \hline
\colhead{Object} & \colhead{Pair} & \colhead{$\Delta m$} }
%&
%\colhead{$R_E/R_{eff}$} & \colhead{\shortstack{Map size \\ in $\eta_0$ }} & 
%\colhead{\shortstack{Pixel size \\ in $\eta_0$ }
\startdata 
HE0230$-$2130	&	B$-$A    &	  0.90 \\
		&	C$-$A    &	 -0.21 \\
		&	D$-$A    &	 -0.76 \\
MGJ0414+0534	&	A2$-$A1  &	   0.56  \\  
		&	B$-$A1   &	 -0.53 \\   
		&	C$-$A1   &	 -0.32 \\
HE0435$-$1223	&	B$-$A    &	  1.14 \\
		&	C$-$A    &	  1.12 \\
		&	D$-$A    &	  0.63 \\
RXJ0911+0551	&	B$-$A    &	  1.74 \\
		&	C$-$A    &	  2.29 \\
		&	D$-$A    &	 0.16 \\
SDSSJ0924+0219	&	B$-$A    &	  0.34 \\
		&	C$-$A    &	  1.27 \\
		&	D$-$A    &	  2.00 \\
PG1115+080	&	A2$-$A1  &	  1.94 \\
		&	B$-$A1   &	 -0.51 \\
		&	C$-$A1   &	 -0.01 \\
RXJ1131$-$1231	&	B$-$A    &	 -3.01 \\
		&	C$-$A    &	 -2.24 \\
		&	D$-$A    &	 -3.43	 \\
SDSSJ1138+0314	&	B$-$A    &	  0.53 \\
		&	C$-$A    &	  0.90 \\
		&	D$-$A    &	 0.90 \\
B1422+231	&	B$-$A    &	  0.84 \\
		&	C$-$A    &	  0.08 \\
		&	D$-$A    &	 -0.16  \\
WFIJ2033-4723	&	A2$-$A1  &	  -0.68 \\
		&	B$-$A1   &	 -0.63 \\
		&	C$-$A1   &	 -0.67 \\
\hline 
\hline
\multicolumn{3}{c}{Optical}\\
\hline
\tablehead{\multicolumn{3}{c}{Optical}\\}
HE0047$-$1756	&	B$-$A	&	-0.19 \\	  	
HE0435$-$1223	&	B$-$A	&	-0.24	 \\  	
		&	C$-$A	&	-0.30	 \\  	
		&	D$-$A	&	 0.09    \\      
SDSS0806+2006	&	B$-$A 	&	-0.47 \\
SBS0909+532	&	B$-$A  	&	-0.60 \\
SDSS0924+0219	&	B$-$A	&	 0.00 \\
FBQ0951+2635	&	B$-$A  	&	-0.69 \\
QSO0957+561	&	B$-$A  	&	-0.30 \\
Q1017$-$20	&	B$-$A  	&	-0.26 \\
HE1104$-$1805	&	B$-$A	&	 0.60  	 \\
PG1115+080	&	A2$-$A1	&	-0.65  	 \\
B1422+231	&	A$-$B	&	 0.16  \\	
		&	C$-$B	&	0.02 	 \\
		&	D$-$B	&	-0.08  	 \\
SBS1520+530	&	B$-$A  	&	-0.39 \\
WFIJ2033-4723	&	B$-$C	&	-0.50 	 \\	
		&	A2$-$A1	&	 0.00 \\
\enddata
\end{deluxetable}

The source brightness is modelled as a Gaussian $I(r)\propto \exp(-r^2/2r_s^2)$. Mortonson et al. (2005) show that 
estimates of the half light radius $R_{1/2}$ depend little on
the specific shape of the radial profile. For a Gaussian, the half light radius is $R_{1/2}=1.18 r_s$. We convolve the 
magnification maps with Gaussians of 16 different sizes over a logarithmic
grid $\ln(r_s/0.05\, \mbox{lt-days})=0.3\times k$ with $k=0,\cdots, 15$, which spans
$r_s\sim 0.05$ to $r_s\sim 4.5$ light-days. 

\begin{figure}
\epsscale{0.85}
\plotone{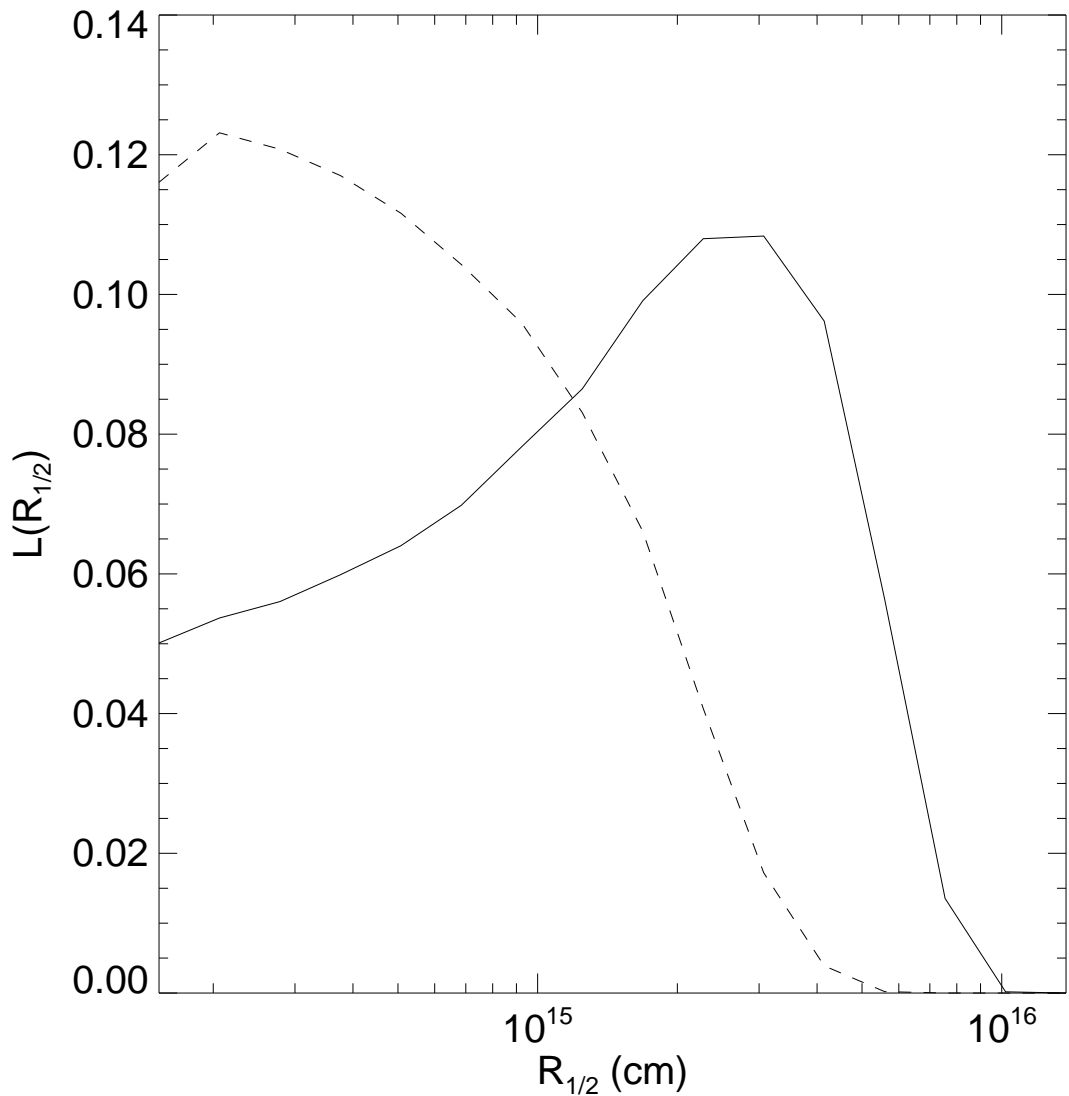}
\caption{Likelihood function for the size of the X-ray emission region $R_{1/2}$ for microlenses of mass $M=0.3M_\sun$. The dashed line shows the joint likelihood for the whole sample. The continuous line shows the likelihood excluding RXJ~1131$-$2131. \label{fig1}}
\end{figure}

We want to compare the observed X-ray fluxes of the images with the predictions of a microlensing model as a function of $r_s$.
We model the magnitude of image $i$ as
\begin{equation}
\label{eq1}
m_i=m_0+\mu_i+\Delta m_i
\end{equation}
where $\mu_i$ and $\Delta m_i$ are, respectively, the macro and micro magnifications of image $i$. As we do not know the intrinsic flux of the source $m_0$, we will use one of the other three images as a reference.
We can calculate the differential microlensing magnification between image $i$ and the reference $r$ as
\begin{equation}
\label{eq2}
\Delta m_{ir}=m_i-m_r-(\mu_i-\mu_r)=(\Delta m_i-\Delta m_r),
\end{equation}
where the difference in the macro magnifications for each image can be accounted for from the lens model. The X-ray microlensing magnifications are presented in Table \ref{tab1}. In principle, errors in the macro model or other secondary effects (e.g. millilensing, extinction, intrinsic variability) can introduce additional noise in our differential microlensing estimates. The possible influence of these effects has been thoroughly analyzed by Schechter et al. (2014), who found that they have a modest influence in their microlensing results, with the
largest uncertainty coming from possible errors in the macro model. We try to account for this through the assumed uncertainty $\sigma=0.2$ mags in the differential magnifications. We can compare the measured differential microlensing magnitude with the prediction of the model and calculate a likelihood
for parameter $r_s$ using the observed differential microlensing magnifications, $\Delta m_{ir}^{obs}$,
\begin{equation}
\label{likel1}
 L(r_s|\Delta m_{ir})=p(\Delta m_{ir}|r_s)=\sum_i \sum_r e^{-\chi^2/2}, 
\label{lik1}
\end{equation}
where 
\begin{equation}
\chi^2=\frac{(\Delta m_{ir}-\Delta m_{ir}^{obs})^2}{\sigma^2}
\end{equation}
and $\sigma$ is a typical error in the estimate of $\Delta m_{ir}^{obs}$ which we have taken as 0.2 mags. 
The summations in Equation \ref{likel1} are over $10^4$ points in the convolved magnification maps of images $i$ and $r$ respectively. The likelihood in Equation \ref{lik1} is therefore calculated using $10^8$ pixel pairs.
We calculate the total likelihood for lens $k$ by combining the likelihoods of the three image pairs relative to the reference image,
\begin{equation} L_k(r_s)=\prod_{i=1,3} L_i(r_s|\Delta m_{ir}), \end{equation}
and the joint probability distribution
\begin{equation} L(r_s)=\prod_{k=1,10} L_k(r_s), \end{equation}
is the product of the individual probabilities for all the lenses in the sample.
The resulting likelihood function using all 10 lenses favours very small sizes, as shown by the dashed line of Figure \ref{fig1}. The
distribution is, however, dominated by a single object, RXJ~1131$-$2131. If we exclude RXJ~1131$-$2131 from the sample, the likelihood function (see Figure 1) is rather different, with a clear maximum at  $\log(R_{1/2}/{\rm cm})=15.5^{+0.3}_{-0.7}$ (68\% confidence interval), indicating that most objects in the sample have sizes close to $\sim$1 light day. This average value may look large at first sight, but it is in agreement with previous measurements (e.g. Mosquera et al. 2013, Morgan et al. 2008). RXJ~1131$-$2131 is an unusual case. 
It has both the largest microlensing magnifications and the smallest estimated black hole mass,
$M_{BH}=6\times10^7 M_\sun$ (Peng et al. 2006), in the sample. 
\begin{figure}
\epsscale{0.85}
\plotone{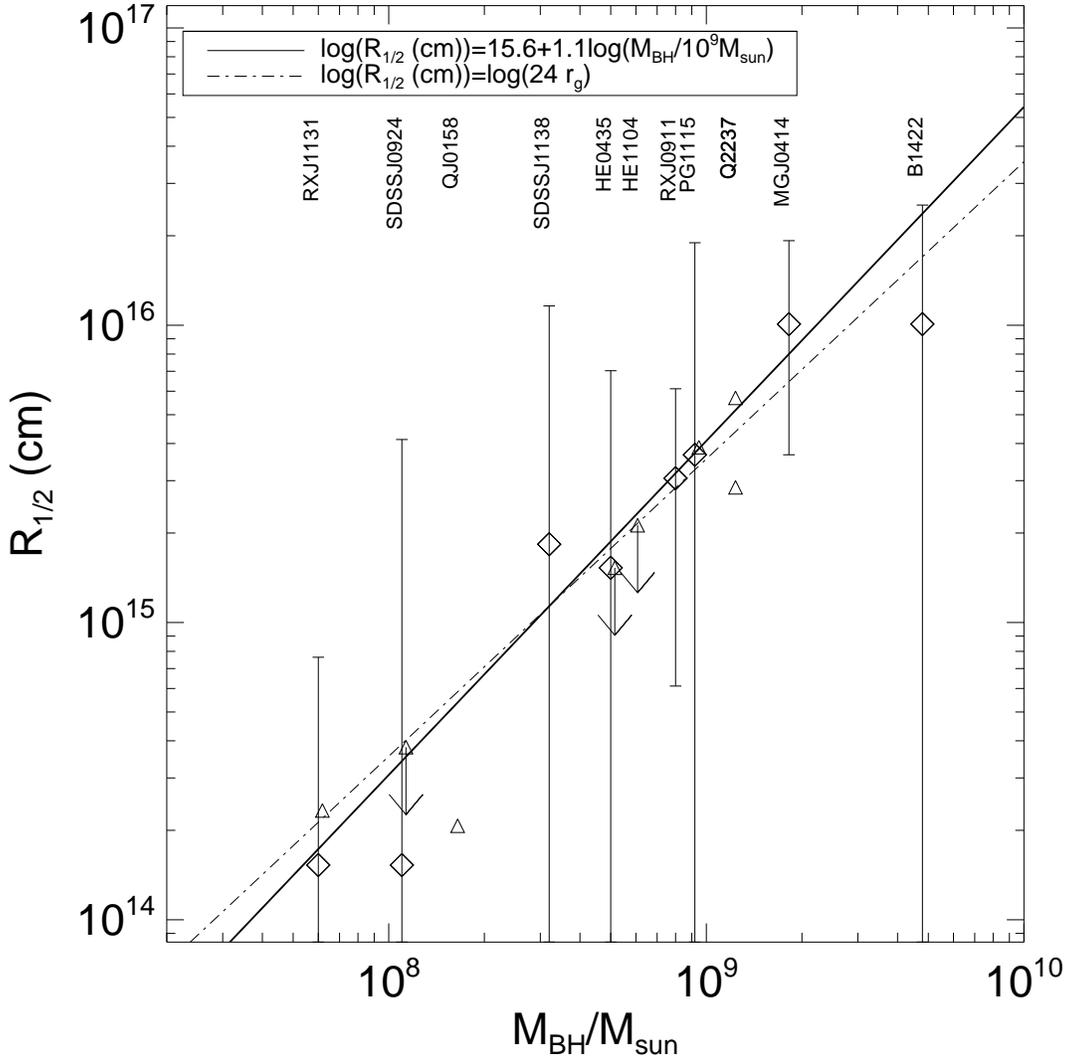}
\caption{Half-light radii, $R_{1/2}$, of the X-ray emission region as a function of the mass of the central black hole $M_{BH}$ (diamonds).
Black hole masses are taken from Peng et al. (2006) 
and Assef et al. (2011). Previous estimates are indicated as triangles for HE0435$-$1223 (Blackburne et al. 2014), HE1104$-$1805 (Blackburne et al. 2015), RXJ~1131$-$2131 (Dai et al. 2010), Q2237+0305 (Mosquera et al. 2013), QJ0158$-$4325 (Morgan et al. 2012), SDSS0924+0219 (McLeod \& Morgan 2014) and PG1115+080 (Morgan et al. 2008). For Q2237+0305, the hard and soft X-ray band estimates are shown separately. Upper limits are indicated by the arrows.  The solid line is a power law fit to the size estimates, $\log(R_{1/2}/{\rm cm})=\log(R_9/{\rm cm})+x\log(M_{BH}/10^9M_\sun)$. The dot-dashed line corresponds to  24 gravitational radii ($R_{1/2}=24r_g=24GM_{BH}/c^2$).\label{fig2}}
\end{figure}
\begin{figure}
\epsscale{0.85}
\plotone{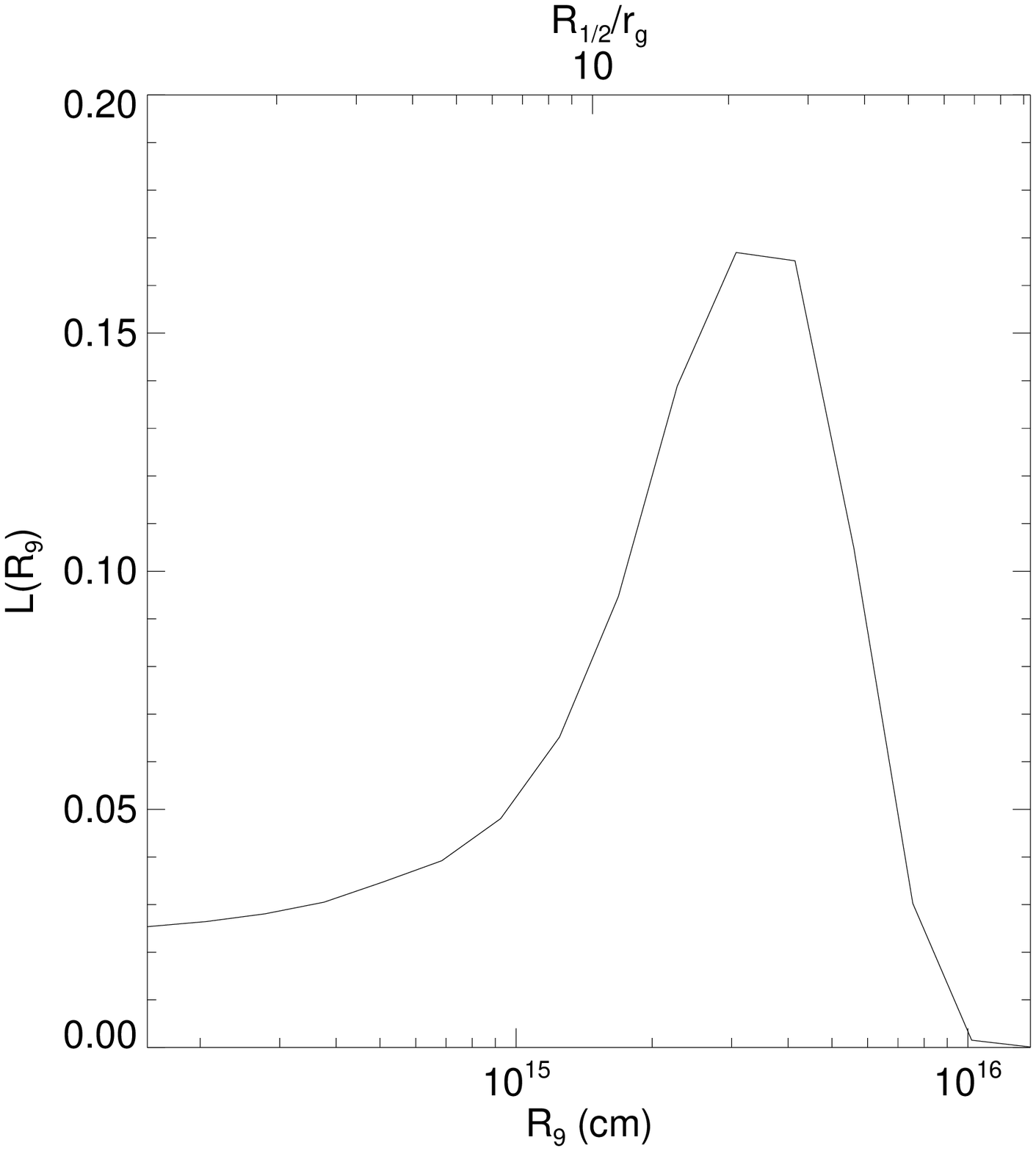}
\caption{Probability distribution for the scaled size of the X-ray emission region size $R_9$ assuming that $R_{1/2}=R_9(M_{BH}/10^9 M_\sun)$.
The upper x-axis shows the size in units of the gravitational radius $r_g=M_{BH}G/c^2$. We assume microlens masses
of $M=0.3M_\sun$. \label{fig3}}
\end{figure}

This suggests that we should examine the scaling with mass even though single epoch microlensing estimates will have large uncertainties.
Figure \ref{fig2} show estimates of the half light radius, $R_{1/2}$, for the eight individual objects in our sample with black hole masses estimates, $M_{BH}$, from Peng et al. (2006) or Assef et al. (2011).
 We also show the X-ray size estimates for HE0435$-$1223 (Blackburne et al. 2014), HE1104$-$1805 (Blackburne et al. 2015), 
RXJ~1131$-$2131 (Dai et al. 2010), Q2237+0305 (Mosquera et al. 2013), QJ0158$-$4325 (Morgan et al. 2012), SDSS0924+0219 (McLeod \& Morgan 2014) and PG1115+080 (Morgan et al. 2008). 
In spite of the large uncertainties (several objects have only upper size limits), there is a clear increase in the size with the mass of the black-hole, and very good agreement with previous estimates from other authors. A fit of our individual size estimates to a 
power law, $\log(R_{1/2}/{\rm cm})=\log(R_9/{\rm cm})+x\log(M_{BH}/10^9M_\sun)$, gives $\log(R_9/{\rm cm})=15.6\pm0.3$, and an exponent $x=1.2\pm 0.5$. The fit can be slightly improved if we include previous size estimates from the literature. In this case, the resulting
paramaters are $\log(R_9/{\rm cm})=15.6\pm0.2$, and an exponent $x=1.1\pm 0.3$. This fit is shown in Figure \ref{fig2} as a continuous line. The exponent is very close to unity, as found by Mosquera et al. (2013), albeit with a large error.
%Irrespective of the exact parameters, Figure \ref{fig2} already shows some clustering of sizes around 1 light-day, showing that it is the typical size of the X-ray emitting region of the lensed quasars. 
In units of the gravitational radius $r_g=M_{BH}G/c^2$, the size of the X-ray emitting region of our sample is reasonably well fit by a line with $R_{1/2}=24r_g$, as also shown in Figure \ref{fig2}.

Taking into account this trend of the X-ray source size with black hole mass, we can now recalculate the joint probability distribution of sizes, but this time we scale the size of each individual object with the mass of its black hole as $R_{1/2}=R_9(M_{BH}/10^9 M_\sun)$. In this case, we will not need to exclude RXJ~1131$-$2131, as its contribution to the joint probability distribution is properly scaled to take account of the low mass of its black hole. In Figure 3 we show the probability distribution for $R_9=R_{1/2}/(M_{BH}/10^9 M_\sun)$. Despite including objects with a wide range in black hole masses, this likelihood is more peaked than Fig 1, which is an indication that scaling the size linearly with the black hole mass has reduced the scatter. 

The maximum likelihood result for the scaled size is $\log(R_9/{\rm cm})=15.5^{+0.2}_{-0.3}$ 
%(or, equivalently, $\log(r_9/{\rm lt-days})=0.0^{+0.2}_{-0.3}$)
, in very good agreement with the results of Mosquera et al. (2013) (their Fig. 4) and with the results of the linear scaling shown in Figure \ref{fig2}. Figure \ref{fig3} also shows
that the size of the X-ray emitting region is restricted to a rather narrow region of $R_{1/2}=(24\pm12)r_g$ in units of the gravitational radius $r_g$. Our present estimate, based on completely independent method and dataset, is in excellent agreement with previous estimates from microlensing (e.g. Mosquera et al., 2013, Blackburne et al., 2014, 2015, Morgan et al., 2012, Dai et al., 2010 ), but also in
good agreement with size estimates by other means such as absorption variability (cf. Agis-Gonz\'alez et al., 2014, Sanfrutos et al., 2013, Uttley et al., 2014). Estimates of the height of the corona above the accretion disk from reverberation lags are also in the range of a few $r_g$ (Reis \& Miller 2013, Emmanoulopoulos 2014, Cackett et al. 2014, Shappee et al. 2014)

\section{Joint Determination of the Stellar Mass Fraction and the X-ray Size}\label{sec3}\

\begin{figure}
\epsscale{0.85}
\plotone{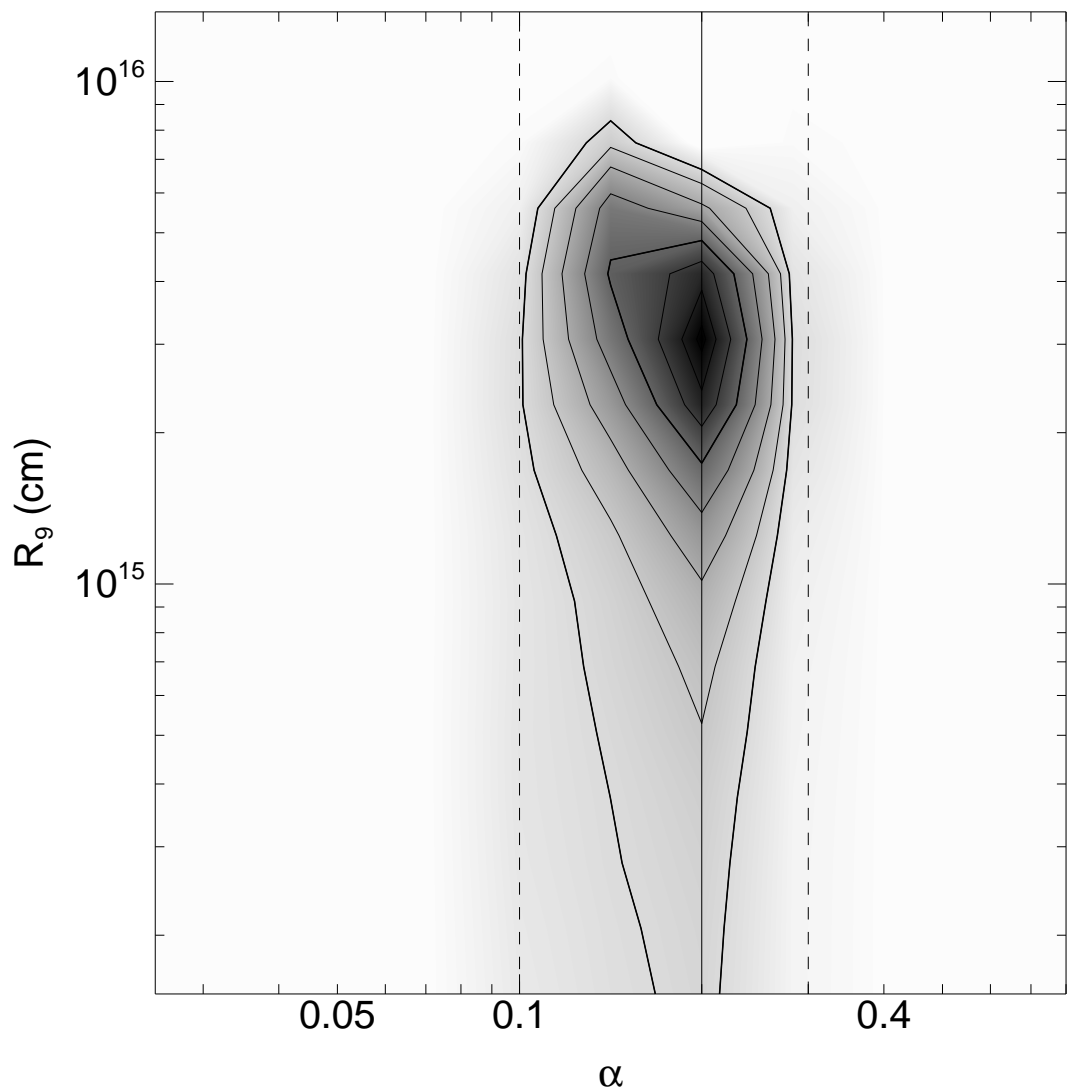}
\caption{Likelihood function for the stellar mass fraction $\alpha$ and the X-ray source size $R_9$ at $M_{BH}=10^9 M_\sun$ for microlenses of mass $M=0.3M_\sun$. The contours are drawn at likelihood intervals of 0.25$\sigma$ for one parameter from the maximum. The contours at 1$\sigma$ and 2$\sigma$ are heavier. Vertical lines indicate the estimate (solid line) and 68\% confidence interval (dashed lines) from a similar analysis of optical microlensing data by Jim\'enez-Vicente et al. (2015). \label{fig4}}
\end{figure}

Next, we allow the stellar mass fraction to vary as well. We simply repeat the calculations, but now include
a logarithmic grid for the stellar mass fraction $\alpha$ such that
$\alpha_j=0.025\times 2^{j/2}$ with $j=0,\cdots,10$, so that $\alpha$ ranges from 0.025 to 0.8.
We again linearly scale the source size $R_{1/2}$ with the black hole mass.
For every image pair, a likelihood is calculated for each of the 176 possible combinations of parameters $(\alpha, R_9)$
to compute
\begin{equation}
L(\alpha,R_9| \Delta m_{ir})=p(\Delta m_{ir}|\alpha,{R_9}).
\end{equation}
From these we can calculate likelihoods for each individual object $L_k(\alpha,R_9)$, and a joint likelihood as
the product of the eight individual likelihoods
\begin{equation}L(\alpha,R_9)=\prod_{k=1,8} L_k(\alpha,R_9). \end{equation}  
The resulting 2D likelihood function is shown in Figure \ref{fig4}.
The joint likelihood distribution shows much less covariance between $\alpha$ and $R_{1/2}$ than found in the optical (cf. Jim\'enez-Vicente et al. 2015). 
In fact, $\alpha$ is very well constrained with little dependence on size, as expected for small sizes. 
The maximum likelihood estimate is $\alpha=0.20\pm0.05$ (for an average distance of $\sim 1.9R_e$), which is in 
excellent agreement with the optical result, as shown in Figure \ref{fig4}.
This agreement strengthens our confidence in the robustness of the method. 
The maximum likelihood estimate for the average size of the emitting region in X-rays is $\log(R_9/{\rm cm})=15.5\pm0.2$, 
%or equivalently $\log(r_9/{\rm lt-day})=0.0\pm0.2$ 
(68\% confidence interval). This estimate is very similar to the obtained in the previous section in a single parameter analysis, which is not surprising, as we used a value of $\alpha=0.2$ which matches the best estimate of the new analysis. Again, the two parameter analysis confirms that the X-ray emitting region is roughly
$R_{1/2}=(24\pm12) r_g$ in size. 
We also recomputed the results sequentially dropping each lens and found that the results are not dominated by any single system.

Taking into account the results of Jim\'enez-Vicente et al. (2015) for the average size of accretion disks in the optical (at an
average rest wavelength of 1736 \AA) of $R_{1/2}^{opt}=7.9^{+3.8}_{-2.6}$ light days, the typical ratio of the half light radius in the optical and X-rays is $R_{1/2}^{opt}/R_{1/2}^{X-ray}\sim 8$, which is in reasonable good agreement with previous results from studies of individual lenses (Morgan et al. 2008, Chartas et al. 2009, Dai et al. 2010, Morgan et al. 2012, Blackburne et al., 2015, 2014, Mosquera et al. 2013, MacLeod \& Morgan 2014). As pointed out by Mosquera et al. (2013), if $R_{1/2}^{X-ray}\propto M_{BH}$ and $R_{1/2}^{opt}\propto M_{BH}^{2/3}$ (cf. Morgan et al. 2010), then $R_{1/2}^{opt}/R_{1/2}^{X-ray}\propto M_{BH}^{-1/3}$, and this ratio should be
larger for smaller masses. Our ratio of $R_{1/2}^{opt}/R_{1/2}^{X-ray}\sim 8$ is a typical value for a mass of $M_{BH}\sim 10^9M_\sun$, which is roughly the average black hole mass in our sample.

\section{The Stellar/Dark Matter Surface Mass Density Profile}

\begin{figure}
\epsscale{0.85}
\plotone{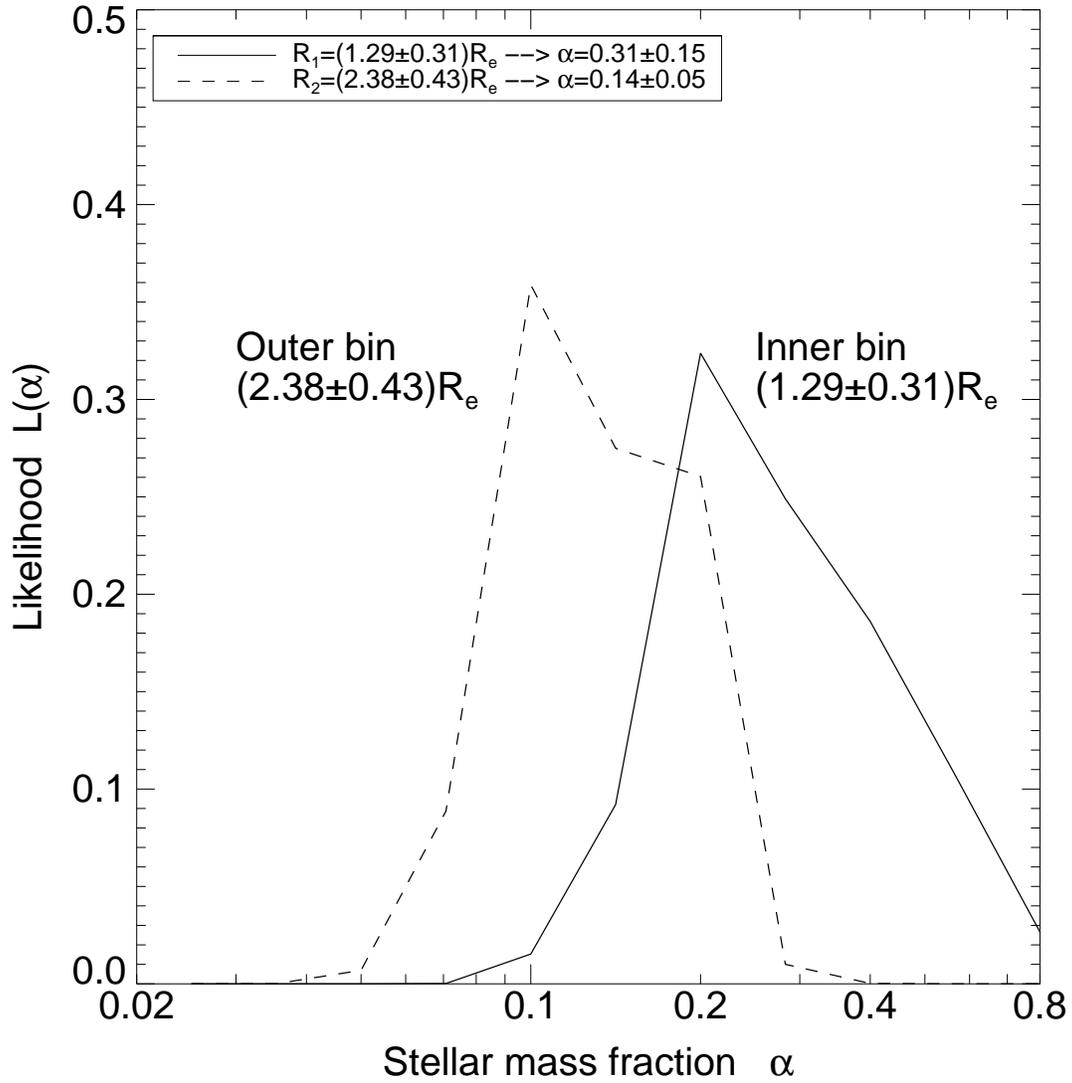}
\caption{Probability distributions for the stellar mass fraction $\alpha$ at two different radii. The continuous (dashed) line is the likelihood function for the inner (outer) radius bin.  
\label{fig5}}
\end{figure}
\begin{figure}
\epsscale{0.85}
\plotone{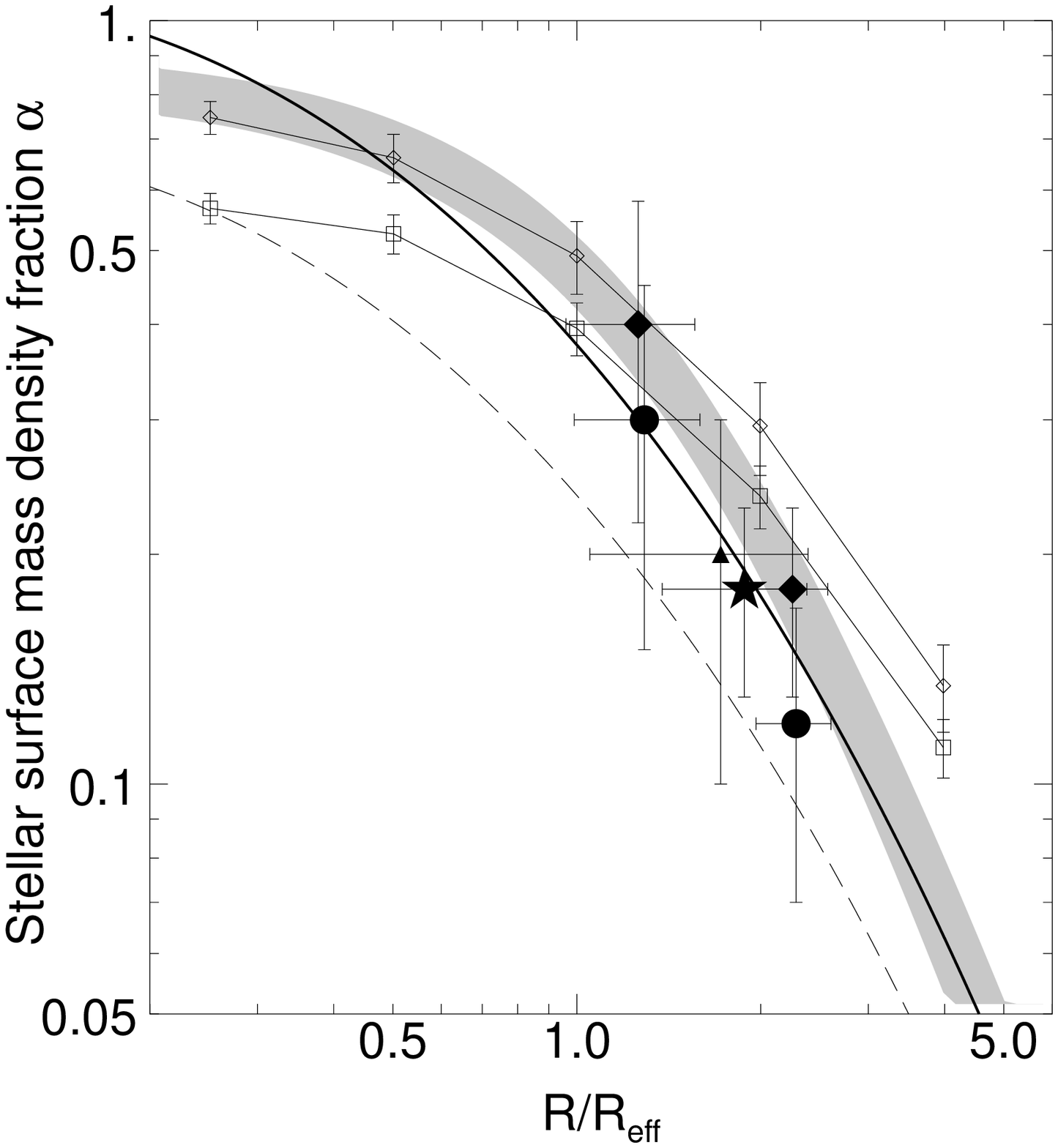}
\caption{Radial profile for the stellar mass fraction. 
The star (triangle) use only the X-ray (optical) data and a single radial bin for all objects.
Circles (diamonds) are the estimates for the two radial bins using a logarithmic (linear) prior on size and both the X-ray and optical data. 
The dashed line corresponds to a simple model with a de Vaucouleurs stellar component and a total mass corresponding to a SIS with a flat rotation curve equal to the maximum rotational velocity of the stellar component. The thick line is the fiducial galaxy model from Schechter et al. (2014), which is a rescaled version of the previous model. The grey band is the best fit profile for the sample of lenses analyzed by Oguri et al. (2014). The open diamonds and squares correspond to a model using a Hernquist component for the stars, embedded in an NFW halo with (open squares) and without (open diamonds) adiabatic contraction of the dark matter, also from Oguri et al. (2014) \label{fig6}}
\end{figure}
Microlensing measurements of the stellar mass fraction (including remnants) have the advantages of being local (not integrated within a certain radius) and insensitive to the stellar IMF. 
On the other hand, the estimates can only be made at the location of the multiple images, which do not sample a broad range of radii (particularly in quadruply imaged systems). 
Moreover, single epoch microlensing provides individual estimates with large uncertainties, making necessary the combination of the estimates from a large enough sample of systems to reduce the uncertainties. 
Here, we combine the X-ray and optical microlensing estimates to examine $\alpha$ (cf. Table \ref{tab1}). With this enlarged sample, we split the data into two radial bins and derive independent estimates for the stellar mass fraction at two different radii $R_E/R_{e}$, where $R_E$ is the Einstein
radius and $R_{e}$ is the effective radius of the lens galaxy. For the X-ray sample we have taken the ratios from Schechter et al. (2014), and for the optical sample, we used the estimates from Jim\'enez-Vicente et al. (2015) for the 18 pairs in the 13 lenses with available estimates of $R_E/R_{e}$. This gives us a total of 48 independent microlensing estimates for 18 different lens systems (there are five lenses in common to both samples). The radial distribution of the observed pairs is bimodal in $R_E/R_{e}$ with a minimum around 1.7, so we split the sample into two radial bins containing objects with $R_E/R_{e}$ smaller and larger than 1.7 respectively. This results in 
average distances of $(1.3\pm0.3) R_{e}$ and $(2.4\pm0.4) R_{e}$ for these two bins, where we are giving the dispersion about the mean
of each bin, not the uncertainty in the mean. We combine the marginalized (using a logarithmic prior on the size $r_s$) probability distributions for all the pairs in each of the two bins to produce a joint probability
distribution for $\alpha$ at these two radii. The result of this procedure is shown in Figure \ref{fig5}.
This is the first direct detection of a significant radial variation of the stellar mass fraction using microlensing measurements. 
The resulting Bayesian estimates (using a logarithmic prior in $\alpha$) for the stellar mass fraction at these radii are $\alpha=0.31\pm0.15$ and $\alpha=0.14\pm0.05$. 

We have also repeated the calculations using a linear prior on size. In this case, the estimates are $\alpha=0.40\pm0.18$ and $\alpha=0.18\pm0.05$ for the inner and outer radial bins respectively.
In Figure \ref{fig6} we also compare these new estimates to our earlier optical microlensing results (Jim\'enez-Vicente et al. 2015), a simple theoretical model and estimates for a galaxy sample based on strong lensing models (Oguri et al. 2014). The simple theoretical model is an early-type galaxy consisting of a de Vaucouleurs component for the stars and a singular isothermal sphere (SIS) for the total mass. Two different scalings of this model are shown. The dashed line is scaled so that the (flat) rotation curve of the SIS equals the maximum rotational velocity of the de Vaucouleurs stellar system. The continuous thick line based on the mass fundamental plane scaling from Schechter et al. (2014), which they take as their fiducial galaxy. Despite the relatively large errors, our results are in good agreement with simple theory and previous results. In particular, there is very good agreement with both the strong lensing results by Oguri et al (2014) and the fiducial galaxy based on Schechter et al. (2014).

\section{Conclusions}

We have performed a statistical analysis of the effect of source size and the fraction of surface mass densisty in stars 
on the microlensing in X-rays for a sample of 10 lensed quasars taken from Schechter et al. (2014). 

\begin{enumerate}

\item Pre-existing studies of X-ray microlensing have found that sizes increase roughly in proportion to the estimated black hole
mass (Mosquera et al. 2013). From a fit of our individual size estimates to a power law $\log(R_{1/2}/{\rm cm})=\log(R_9/{\rm cm})+x\log(M_{BH}/M_\sun)$, we find $\log(R_9/{\rm cm})=15.6\pm0.3$, and $x=1.2\pm 0.5$ (improving slightly when previous size estimates are included to $\log(R_9/{\rm cm})=15.6\pm0.2$, and $x=1.1\pm 0.3$).
Based on this, we assumed a linear scaling ($x=1$) of size $R_{1/2}$ with black hole mass $M_{BH}$, and we find an average size 
for $M_{BH}=10^9M_\sun$ of $\log(R_9/{\rm cm})=15.5^{+0.2}_{-0.3}$.
 This result is rougly consistent with the simple scaling $R_{1/2}\simeq(24\pm12) r_g$ (with $r_g=GM_{BH}/c^2$). This agrees well with previous determinations
for individual sources (e.g. Morgan et al. 2008, Blackburne et al. 2015, Mosquera et al. 2013).

\item Microlensing in X-rays produces an estimate for the local surface density in stars and stellar remnants of $\alpha=0.20\pm0.05$ at a typical radius of 1.9 effective radii, in excellent agreement with the independent result obtained using microlensing in the optical by Jim\'enez-Vicente et al. (2015).
 
\item By combining the microlensing estimates in X-rays and in the optical, we have been able to obtain the stellar mass fraction at two different radii. We find a drop in the stellar mass fraction from $\alpha=0.31\pm0.15$ at a radius of $(1.3\pm0.3)R_{e}$ to $\alpha=0.13\pm0.05$ at $(2.3\pm0.3)R_{e}$. This result is in very good agreement, given the uncertainties, with results from strong lensing analysis of a large sample by Oguri et al. (2014) and with the scaling of the mass fundamental plane found by Schechter et al. (2014). 

\end{enumerate}

The application of the present method to a significantly larger sample of lens systems with measured microlensing (preferably, but not necessarily, in X-rays) should allow the determination of the shape of the dark matter density profile relative to the stars in the radial range from roughly 0.5 and 3 effective radii.
\

\noindent Acknowledgements:

The authors would like to thank M. Oguri for kindly providing the differential version of their results for comparison with the present work shown in Figure \ref{fig6}. 
This research was supported by the Spanish Mi\-nis\-te\-rio de Educaci\'{o}n y Ciencia with the grants AYA2011-24728, AYA2010-21741-C03-01 and AYA2010-21741-C03-02. JJV is also supported by the Junta de Andaluc\'{\i}a
  through the FQM-108 project. JAM is also supported by the Generalitat Valenciana with the grant PROMETEOII/2014/060.
CSK is supported by NSF grant AST-1009756.

\end{document}